\documentstyle[11pt,aaspp4,flushrt]{article} 
\input psfig
\begin{document}
\title{GRB Phenomenology, Shock Dynamo, and the First Magnetic Fields} 
\author{Andrei Gruzinov}
\affil{Physics Department, New York University, 4 Washington Place, New York, NY 10003}

\begin{abstract}
A relativistic collisionless shock propagating into an unmagnetized medium leaves behind a strong large-scale magnetic field. This seems to follow from two assumptions: (i) GRB afterglows are explained by synchrotron emission of a relativistic shock, (ii) magnetic field can't exist on microscopic scales only, it would decay by phase space mixing. Assumption (i) is generally accepted because of an apparent success of the shock synchrotron phenomenological model of GRB afterglow. Assumption (ii) is confirmed in this work by a low-dimensional numerical simulation. One may hypothesize that relativistic shock velocities are not essential for the magnetic field generation, and that all collisionless shocks propagating into an unmagnetized medium generate strong large-scale magnetic fields. If this hypothesis is true, the first cosmical magnetic fields could have been generated in shocks of the first virialized objects. 

\end{abstract}

\section{GRB Afterglow Phenomenology}

It appears that GRBs and GRB afterglows can be adequately described by synchrotron emission of relativistic shocks (eg. Waxman 1997, Sari et al 1997, Guetta et al 2000). These phenomenological models describe the shock by a set of adjustable parameters. One of these parameters, $\xi _B\equiv B^2/(8\pi \epsilon )$, is the magnetic energy density divided by the proton energy density $\epsilon$ downstream of the shock transition. It is thought that $\xi _B$ is close to unity, say $\xi _B\sim 0.1$ . Much smaller values, $\xi _B\sim 0.001$, were given by Panaitescu \& Kumar (2000), but as we are about to show even 0.001 is a surprisingly large number.

For the sake of simplicity, consider a mildly relativistic stage of a GRB afterglow, when Lorentz factor is $\sim$few. Assume that unshocked plasma is at a typical ISM temperature, $\sim 1eV$. The unshocked magnetic energy per particle is $\lesssim 1eV$, because the field has to be confined by the plasma pressure. If hydrodynamics is applicable, then the blast wave is stable (Gruzinov 2000), and there is no magnetic field generation. Magnetic field is amplified only by compression. Then the shocked  magnetic energy per particle is $\lesssim$ few eV, because the shock has a compression factor $\sim$few. But the proton temperature in the shocked plasma is $\sim$ few GeV, giving $\xi _B\sim 10^{-9}$, which is at least 6 orders of magnitude smaller than required by the synchrotron phenomenological model of GRB afterglow.

In this work, we assume that the synchrotron GRB model is correct. Then the magnetic field should be somehow generated  by the shock. For all practical purposes, the unshocked plasma is unmagnetized. The nature of this shock-generated magnetic field is an interesting unsolved problem. Here we try to guess an answer, but we have to acknowledge the speculative nature of our discussion. The right way to solve this problem is direct numerical simulations of a three dimensional collisionless (Vlasov) plasma.  

\section{Shock Dynamo}

That collisionless shocks can generate strong, that is near equipartition, magnetic fields is well known (eg. Sagdeev, 1966). Low-dimensional numerical simulations illustrating this process are also available (eg. Kazimura et al 1998, Appendix). The mechanism of the magnetic field generation is the Weibel instability, which can be explained as follows. Within the shock transition region, the distribution function of charged particles is strongly anisotropic. Particles move predominantly perpendicular to the front. These particles might be thought of as straight currents. Currents of the same sign attract each other, leading to bunching of electrons moving in one direction and protons moving in the opposite direction. This current bunching generates magnetic fields.

The length and time scales characterizing the instability can be obtained from dimensional analysis. The instability is collisionless --- it is described by the Vlasov system of equations (that is collisionless Boltzmann plus Maxwell). The Vlasov system contains the following dimensional quantities, $nm$, $ne$, and $c$. Here $n$ is the unshocked plasma density, $m$ is the proton mass, $c$ is the speed of light.\footnote{For the sake of simplicity, we will exclude a small dimensionless number $m_e/m$, where $m_e$ is the mass of electron, from the list of parameters. In reality this number might be important, for example, Sagdeev (1966) argues that non-relativistic Weibel saturates at $\xi _B\sim m_e/m$.} We will assume that the shock is mildly relativistic, so it is not characterized by any large or small dimensionless numbers. Then there is only one time scale in the problem, the inverse plasma frequency, $\omega _p^{-1}$, where $\omega _p\equiv (4\pi n e^2/m)^{1/2}$. The only length scale is the skin depth, $\delta \equiv c/\omega _p$. 

Now the initial stage of the Weibel instability might be described as follows. The shock width is $\sim \delta$. Within the shock, a near equipartition magnetic field is generated. The field isotropizes the distribution functions of charged particles. The characteristic length scale of the generated field is $\sim \delta$.

But what happens many deltas downstream (Gruzinov and Waxman, 1998)?  The standard plasma-physical answer would be: (i) Strong electromagnetic fields occupy a layer several $\delta$ wide. Within this layer, electromagnetic fields play a role of collisionality, bringing the plasma up the shock adiabatic. Many deltas downstream, there is only shocked plasma and virtually no fields. Electromagnetic fields are week, because they decay by phase space mixing (Landau damping). We have confirmed this magnetic field decay scenario, which is the common collisionless shock wisdom, by two-dimensional numerical simulations (Appendix). 

But: the decaying field scenario is ruled out by astronomical observations, so long as we believe that GRBs and GRB afterglows are synchrotron emitting shocks. Synchrotron emission of a few-skin-deep layer would be negligible.

The two most natural possibilities consistent with the synchrotron GRB model are: (ii) strong small-scale fields (scale $\sim \delta$) exist downstream, (iii) strong large-scale fields (scale $\sim l$, $l$ is proper distance from the shock) exist downstream. Now (ii) is, in fact, impossible. Magnetic fields don't live on length scales $\delta$. They can be born there, and then they decay there on a time scale $\sim \omega _p^{-1}$ due to phase space mixing. Our numerical simulation confirms that magnetic fields really die out on skin depth scales (Appendix), at least in 2D. We will assume that the decay of microscopic, skin-depth magnetic fields by the phase space mixing is a true phenomenon, not an artifact of two dimensions. In 3D the phase space mixing should be even easier. 

Then we are left with the possibility (iii) -- strong (but maybe $\xi _B\sim m_e/m_p$) large-scale (length scale of order distance from the shock) magnetic fields are somehow generated by a collisionless shock propagating into an initially unmagnetized plasma. We do not understand how this happens. We suspect that particle acceleration is an essential part of the process.\footnote{In 2D, we see no magnetic field generation on large scales and no particle acceleration.} The presence of an energetically important population of accelerated particles invalidates the MHD approximation. And one does have to go beyond MHD, because MHD shocks are stable and therefore do not generate magnetic fields.

We think that the problem of magnetic field generation in shocks will be solved by direct numerical simulations. Our amateur numerical program seems to be sufficient for a 2D case. A carefully designed code should solve the 3D problem. The advocated 3D simulation of a collisionless shock propagating into an unmagnetized medium should be performed regardless of our arguments.\footnote{Such simulation is needed to understand particle acceleration. Postulating ``scattering clouds'' is not good enough.} This simulation might prove our scenario wrong, but any result should be of great interest. Say, one discovers that magnetic fields are not produced. Then, what is the nature of the GRB afterglows?  

\section{The First Magnetic Fields}
We have suggested that relativistic collisionless shocks generate strong large scale magnetic fields. However we do not see how the relativistic effects might be important. We therefore hypothesize that non-relativistic shocks also generate strong large scale magnetic fields.

This means, in particular, that the first cosmic magnetic fields could appear together with the first virialized objects. The magnetic fields are generated in virialization shocks, and they are born strong and large-scale. The primordial magnetic field theories are not needed to explain ``the seed fields'' for galactic dynamo.

\begin{appendix}

\section{Weibel instability in 2D3V }
Here we show that magnetic field generated by a 2D3V (two-dimensional in space and three-dimensional in velocity space) Weibel instability decays due to  phase space mixing after a few plasma times. We perform a simplified numerical simulation to illustrate this effect. 

We use dimensionless units $e=m=c=1$. Electromagnetic field is 
\begin{equation}
{\bf E}=(0,0,-\partial _tA), ~~~{\bf B}=(\partial _yA, -\partial _xA,0),
\end{equation}
where the z-component of the vector potential is $A=A(x,y,t)$. The absence of the electrostatic field is explained below. Non-relativistic charges move in a 2D space, ${\bf r}=(x,y)$, with 3D velocities $({\bf v},u)=(v_x,v_y,u)$. This means that charges are rigid rods elongated along the z-coordinate. The equations of motion for positive charges are
\begin{equation}
\dot{{\bf r}}={\bf v},~~~\dot{{\bf v}}=u\nabla A,~~~\dot{u}=-\partial _tA-{\bf v}\cdot \nabla A.
\end{equation}

To simplify our basic system, we will assume that at t=0, and therefore at all later times, each positive charge $({\bf r}, {\bf v}, u)$ is matched by a negative charge $({\bf r}, {\bf v}, -u)$. Under this symmetry, there is no electrostatic component of the field. The only field equation is 
\begin{equation}
-\nabla ^2A+\partial _t^2A=j,~~~j=\Omega^2\int d^2vdu~fu.
\end{equation}
Here $\Omega ^2=8\pi ne^2/m$, $n$ is the density (the unperturbed distribution function $f=F({\bf v},u)$ is normalized $\int d^2vdu~F=1$).

\subsection{Linear Theory}
To analyze the linear Weibel instability, we follow the standard Vlasov route (eg. Lifshitz \& Pitaevskii 1981). We represent the equations of motion by the Boltzmann equation:
\begin{equation}
\partial _tf+{\bf v}\cdot \partial _{\bf r}f+u\nabla A\cdot \partial _{\bf v}f-(\partial _tA+{\bf v}\cdot \nabla A)\partial _uf=0.
\end{equation}
For a perturbation $\propto e^{-i\omega t+i{\bf k}\cdot {\bf r}}$, the perturbed part of the distribution function is 
\begin{equation}
\delta f=A(~(\omega -{\bf k}\cdot {\bf v})^{-1}u{\bf k}\cdot \partial _{\bf v}+\partial _u~)F.
\end{equation}

We use this in (A3), and after integrating the current density by parts, we get the dispersion law (that is relation between $\omega$ and ${\bf k}$):
\begin{equation}
{\omega ^2-k^2\over \Omega ^2}=1+<{u^2\over (\omega -{\bf k}\cdot {\bf v})^2}>k^2,
\end{equation}
where $<...>$ is the average over the unperturbed distribution function.

If, as in our numerical simulation, at the initial time all velocities are parallel to z, (A6) gives the instability growth rate 
\begin{equation}
-\omega ^2\approx {<u^2>k^2\over 1+k^2/\Omega ^2}.
\end{equation}
This means that the fastest growing modes are small-scale, with wavelength  $\lambda \lesssim \Omega ^{-1}$. The characteristic growth rate is $\sim u\Omega$.

\subsection{Nonlinear Theory}

As the instability develops, magnetic field turns the particles' trajectories, and a non-zero velocity dispersion in the (x,y) plane appears. This  stabilizes the short wavelengths first, because in the first non-vanishing order in $v$, (A6) gives, for $k\gg \Omega$,
\begin{equation}
-\omega ^2\approx \Omega ^2<u^2>~-~{3\over 2}{<u^2v^2>\over <u^2>}k^2.
\end{equation}

One expects maximal convertion of kinetic into magnetic energy at about the time when one Larmor circle in the generated magnetic field has been completed by a typical particle. This should occur at length scales somewhat larger than $1/\Omega$, because the generated field has to be somewhat subequipartition, that is less than $\Omega u$.

Further evolution is governed by conservation of the mean squared potential. As follows from translation invariance along z, the generalized momentum, $p=u(t)+A(x(t),y(t))$ is conserved for each particle. In particular, $I=\sum p^2/N$, where the sum is over all $N$ particles, is conserved. $I$ has dimensions of the squared potential, that is $B^2L^2$, where $B$ is the magnetic field and $L$ is the length. One may then suggest the following scenario for the magnetic field evolution. The length scale $L$ grows, $I$ is conserved, therefore the magnetic energy $<B^2>$ decreases as $I/L^2$. To determine $L$ as a function of the evolution time $t$, we assume that $L/t$ is close to the Alfven velocity, which is $\propto B$. It follows that $<B^2>\propto t^{-1}$.

\subsection{Numerical Simulations}

In our simulations, we integrated the particle trajectories (A2), and solved the electrodynamics equations $\partial _tA=-E$, $\partial _tE=-\nabla ^2A-j$ on a grid. Space was a periodic square of size 1. We used the simplest straightforward numerical algorithm. With a 100x100 grid for an electromagnetic field, we needed N=2,000,000 particles to compensate for the discreetness effects. For $\Omega =100$, with the time step of 0.0005, energy is conserved to 0.3\% , and the mean squared potential (see \S A.2) is conserved to 3\%.

Initial conditions are as follows. Electromagnetic field is zero. Particles are at random spatial positions, with ${\bf v}=0$, and $u$ randomly chosen from an interval $(-0.2,0.2)$. This velocity might seem to be too high for a non-relativistic approximation to apply, since in our units the speed of light is 1. However, as we have checked, even the fastest particle in our run does not get faster than about 0.5. Results are shown in figures 1-3. 

At $t=1$, 16\% of energy is converted into magnetic energy. The mean magnetic field is $B\sim 3$. Since velocity is $v\sim 0.1$, the corresponding Larmor radius is $\sim 0.03$, close to the length scale seen in Fig.1. As seen from Fig.2, the length scale of the field grows. As we have said, this is accompanied by the magnetic energy decay. As seen from Fig.3, magnetic energy is indeed described by the $t^{-1}$ law.

\begin{figure}[htb]
\psfig{figure=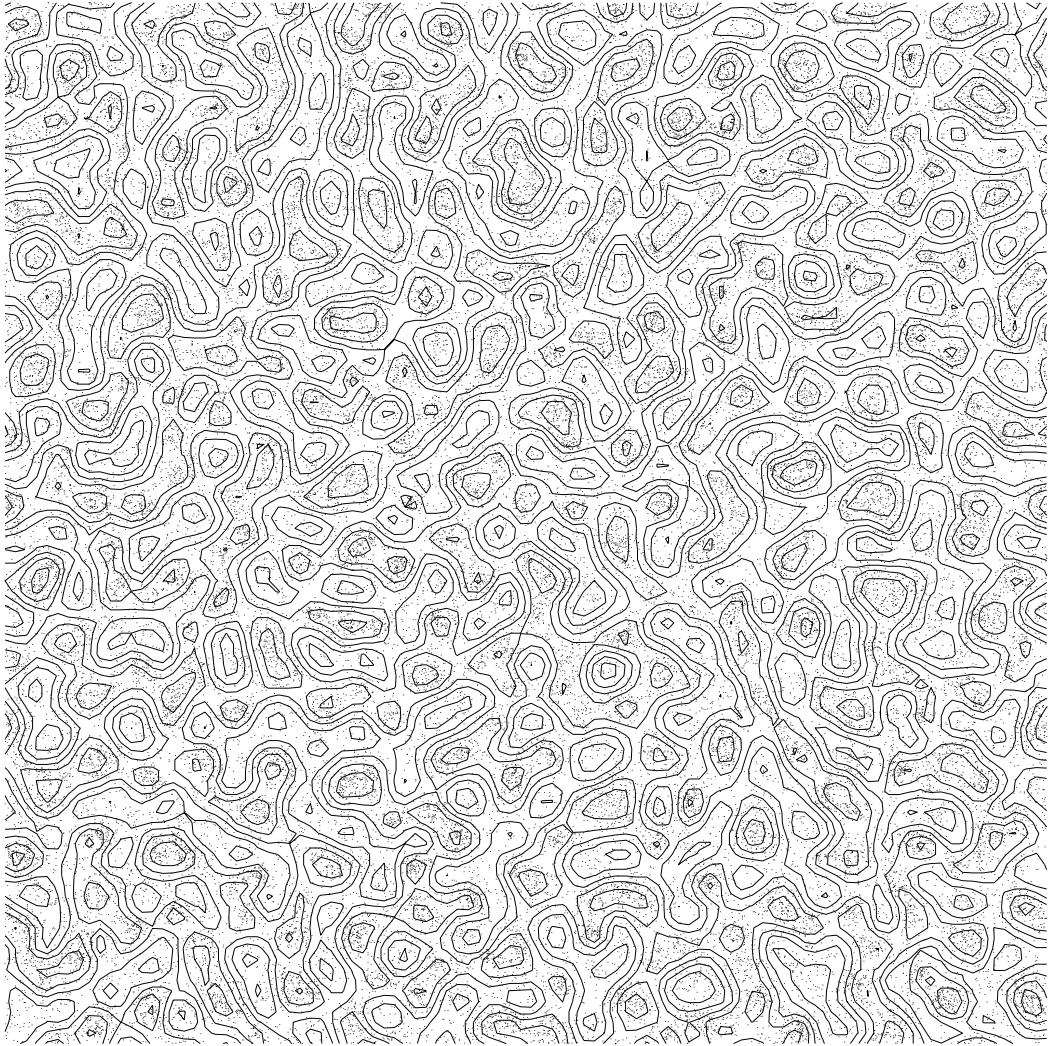,width=6.5in}
\caption{Generation of the magnetic field. Particles with positive u-velocities and the magnetic field at $t=1$, when the field energy is 16\% of the total energy. }
\end{figure}

\begin{figure}[htb]
\psfig{figure=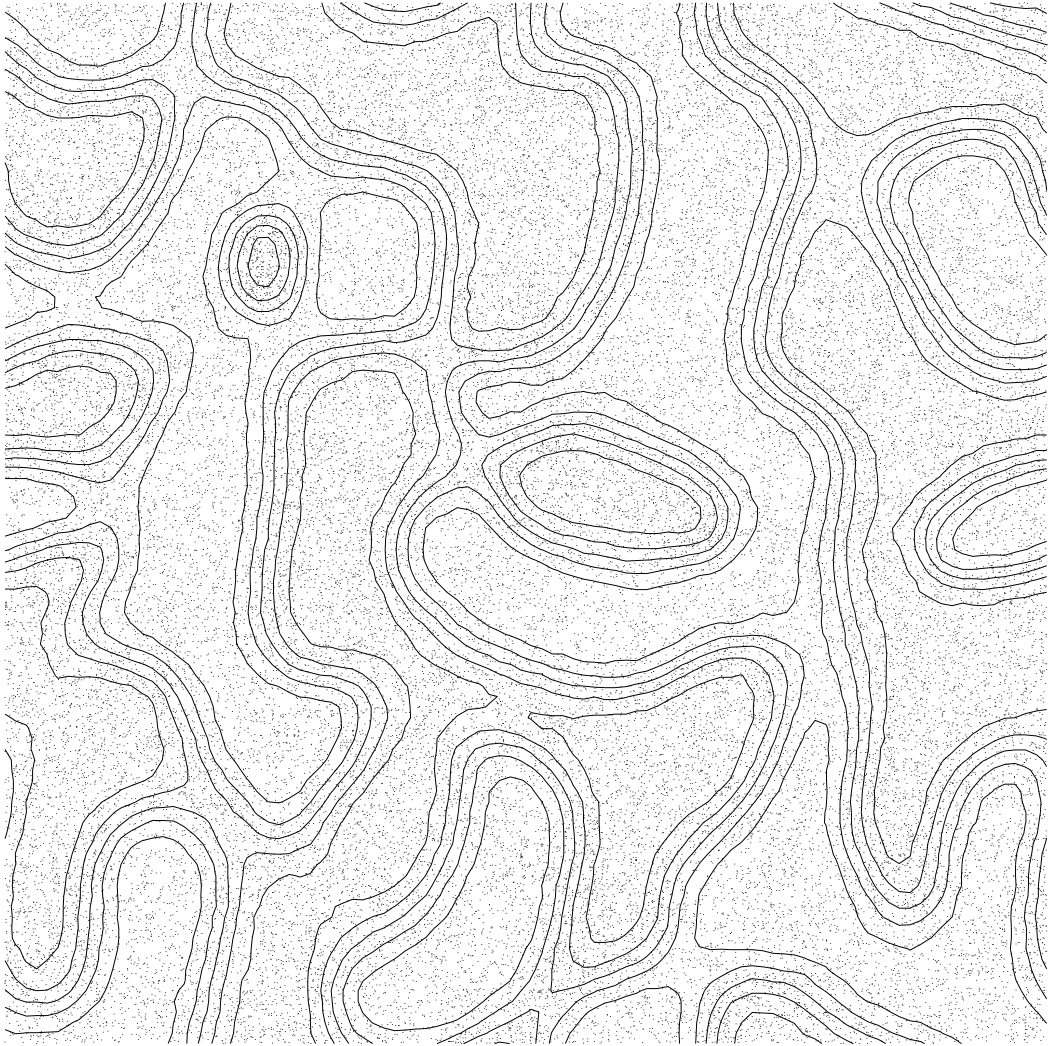,width=6.5in}
\caption{Decay of the magnetic field. Particles with positive u-velocities and the magnetic field at $t=10$, when the field energy is 4\% of the total energy. }
\end{figure}

\begin{figure}[htb]
\psfig{figure=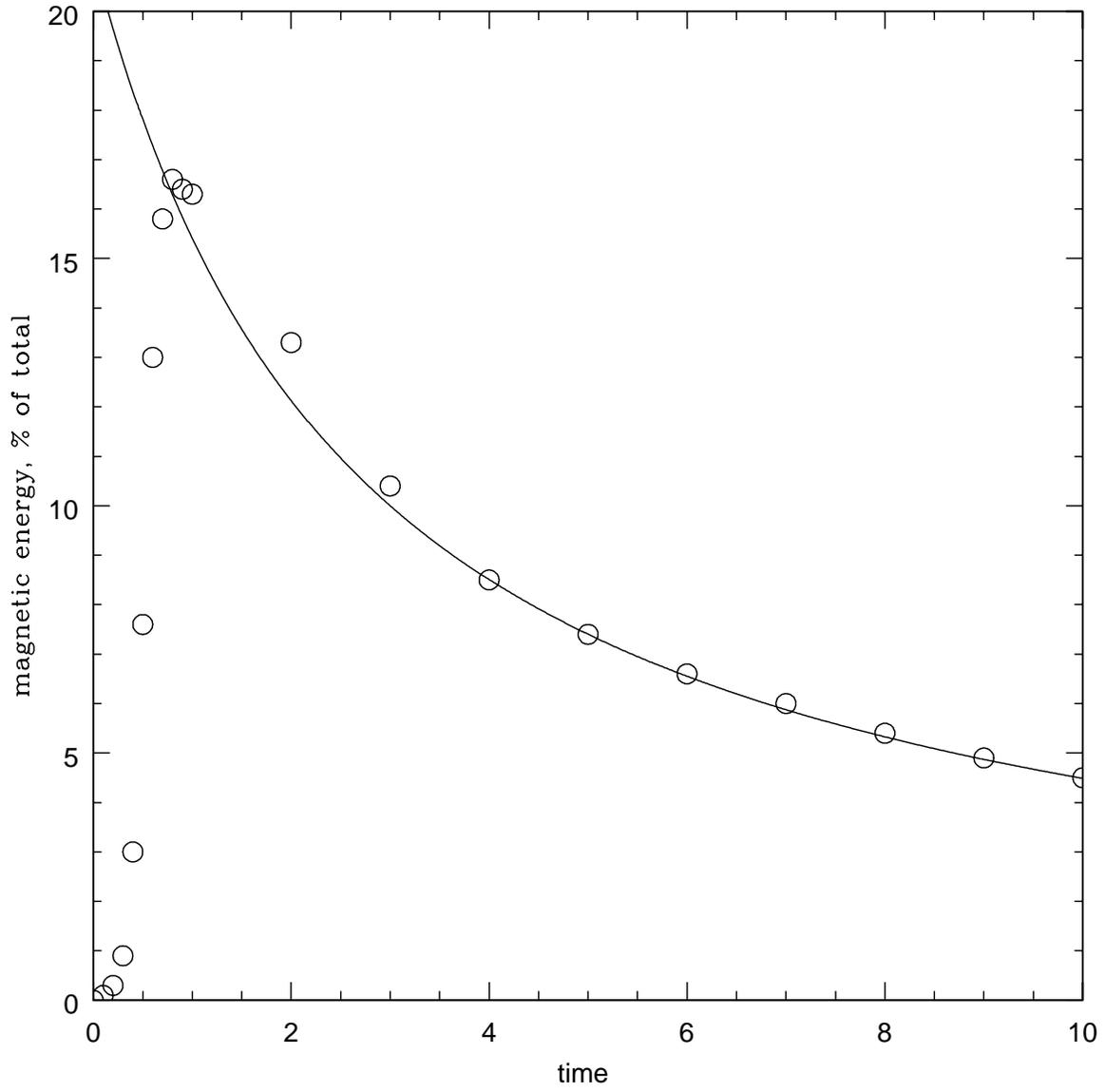,width=6.5in}
\caption{Growth and decay of magnetic energy. Circles: numerical data. Solid line: a two-parameter fitted $t^{-1}$ law: $0.57(2.7+t)^{-1}$. }
\end{figure}

\end{appendix}

\end{document}